\documentclass[%
 twocolumn,
 reprint,
showpacs,
 amsmath,amssymb,
 aps,
]{revtex4}

\usepackage{graphicx}% Include figure files
\usepackage{dcolumn}% Align table columns on decimal point
\usepackage{bm}% bold math
\begin{document}

%\preprint{APS/123-QED\author{Takahiro Ohgoe}}

\title{Novel Mechanism of Supersolid of Ultracold Polar Molecules in Optical Lattices}% Force line breaks with \\

\author{Takahiro Ohgoe$^1$}
\author{Takafumi Suzuki$^2$}
\author{Naoki Kawashima$^1$}%
\affiliation{%
	$^1$Institute for Solid State Physics, University of Tokyo, 5-1-5 Kashiwa-no-ha, Kashiwa, Chiba 277-8581, Japan\\
	$^2$Research Center for Nano-Micro Structure Science and Engineering, Graduate school of Engineering, University of Hyogo, Himeji, Hyogo 671-2280, Japan
}%

\date{\today}% It is always \today, today,

\begin{abstract}
We study the checkerboard supersolid of the hard-core Bose-Hubbard model with the dipole-dipole interaction. This supersolid is different from all other supersolids found in lattice models in the sense that {\it superflow paths} through which interstitials or vacancies can hop freely are absent in the crystal. By focusing on repulsive interactions between interstitials, we reveal that the long-range tail of the dipole-dipole interaction have the role of increasing the energy cost of domain wall formations. This effect produces the supersolid by the second-order hopping process of defects. We also perform exact quantum Monte Carlo simulations and observe a novel double peak structure in the momentum distribution of bosons, which is a clear evidence for supersolid. This can be measured by the time-of-flight experiment in optical lattice systems. 

%PACS numbers: May be entered using the \verb+\pacs{#1}+ command.
\end{abstract}

\maketitle

%\tableofcontents

%\section{\label{sec:level1}Introduction}	
	Recently, there are great experimental efforts to create ultracold dipolar molecules in optical lattices\cite{sage2005, ni2008, ospelkaus2008}. In these systems, owing to the dipole-dipole interaction, various quantum phases are predicted theoretically, such as supersolid, devils staircase, crystals with long periodicities and so on\cite{goral2002, yi2007, lahaye2009, capogrosso2010-1, pollet2010}. Experimentally, the realization of supersolids has attracted great interest. In recent study by quantum Monte Carlo calculations\cite{capogrosso2010-1}, several characteristic supersolid states, such as checkerboard-type and star-type supersolids, were found in hardcore bosonic systems with the dipole-dipole interaction on a square lattice. Since optical lattice is one of ideal tools that can be compared with quantum Monte Carlo results directly, this system is a promising candidate for the realization of supersolids.

	Although the detection of the checkerboard supersolid in hard-core bosonic systems is highly plausible in ultracold polar molecules, its mechanism is not clear. The possibility of supersolid state itself has been already discussed by quantum Monte Carlo calculations for bosonic systems on several lattices such as triangular lattice\cite{wessel2005, boninsegni2005, pollet2010}, face centered lattice\cite{suzuki2007}, and square lattice\cite{batrouni2000, sengupta2005, dang2008}. The mechanism for all these supersolids can be explained by the existence of {\it superflow paths} through which vacancies or interstitials can hop freely in crystals, thus Bose-condensing. In contrast to them, the checkerboard supersolid apparently has no such superflow paths in the hard-core boson case, because the particle hopping is restricted to the nearest-neighbor sites. In fact, it has been established that no supersolid phase appears when the system has only the nearest-neighbor and the next-nearest-neighbor repulsions on the square lattice\cite{dang2008}. Therefore, the long -range tail of the dipole-dipole interaction seems to play an important role to stabilize the supersolid state. However, its role has not been fully discussed yet.
	
	In this paper, we study the mechanism of the checkerboard supersolid in dipolar hard-core bosons in 2D optical lattices. By discussing the stability of the supersolid against the domain wall formation, we reveal that the long-range tail has the effect of increasing the energy cost of the domain wall formations and change the mechanism of the supersolid. We also perform the exact quantum Monte Carlo simulations to investigate the properties of this novel supersolid. As a result, we observe the two types of peaks in the momentum distribution of bosons, which indicate superfluidity and solidity respectively.

%\section{\label{sec:level2}Model}
	We consider the hardcore Bose-Hubbard model with the dipole-dipole interaction on a square lattice. This system can be realized by bosonic polar molecules in a 2D optical lattice. In this paper, electric dipole-dipole moments are assumed to be arranged perpendicular to the 2D plane by applying the electric field. Thus the dipole-dipole interaction is expressed as $1/r^3$ here. The Hamiltonian is given by
	\begin{eqnarray}
		H = - t \sum_{\langle i, j\rangle} ( b_{i}^{\dagger} b_{j} + h.c. ) + V \sum_{i<j} \frac{n_i n_j}{r_{ij}^3} - \mu \sum_{i} n_{i}, \label{eq:hamiltonian}
	\end{eqnarray}
	where $b_{i} (b_{i}^{\dagger})$ is the annihilation(creation) operator of hardcore bosons on the site $i$, $n_{i}=b_{i}^{\dagger} b_{i}$ is a particle number operator. $t$, $V$, and $\mu$ indicate the hopping parameter, the strength of the dipole-dipole interaction, and the chemical potential respectively. $r_{ij}$ is the distance between the site $i$ and $j$. In what follows, we take the lattice spacing as the unit of distance. $\langle i,j \rangle$ means nearest-neighbor pairs. The system size is defined by $N=L \times L$ and the periodic boundary condition is imposed. In this model, the existence of the checkerboard supersolid phase is shown by the quantum Monte Carlo simulations\cite{capogrosso2010-1}.

\begin{figure}[t]
		\includegraphics[width=9cm]{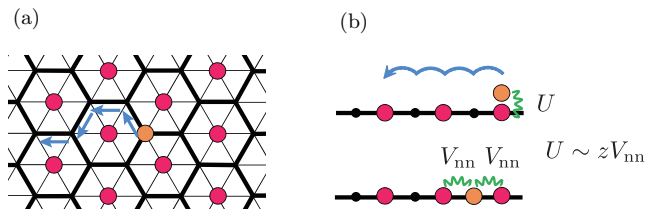}
		\caption{(Color online) Examples of superflow paths for interstitials in crystals. The red circles and the orange circles represent the bosons forming the crystal and interstitial bosons respectively. The paths are denoted by thick lines. The arrows indicate the hopping process of interstitials. (a) The filling factor $\rho=1/3$ crystal of hard-core bosons with the nearest-neighbor repulsion on a triangular lattice. (b) The checkerboard crystal of soft-core bosons with the on-site repulsion $U$ and the nearest-neighbor repulsion $V_{\rm nn}$ on a square lattice. For simplicity, the system is depicted along the one-dimensional direction. The wavy lines represent the repulsive interactions between the interstitial and the checkerboard background.}	
		\label{fig:superflowpath}
	\end{figure}

%\section{\label{sec:level3}Mechanism of supersolid}
    Before we discuss the mechanism of the checkerboard supersolid in the model (\ref{eq:hamiltonian}), we explain that this supersolid is different from previous ones which have been found in the other lattice models by exact quantum Monte Carlo simulations\cite{wessel2005, boninsegni2005, suzuki2007, batrouni2000, sengupta2005, dang2008, pollet2010}. All previous supersolids are related to the presence of superflow paths in crystals. Here, we define superflow paths as paths on which defects such as interstitials or holes in crystals can hop strictly or approximately by the first-order perturbation of $t$. Through these paths, defects in crystal can hop freely and Bose-condense. In Fig. \ref{fig:superflowpath}, we show the examples of superflow paths. If we consider hard-core bosons on a triangular lattice with the nearest-neighbor repulsion, the bosons form a superlattice like Fig. \ref{fig:superflowpath} (a) at the 1/3-filling \cite{wessel2005, boninsegni2005}. In this case, the superflow path is the honeycomb lattice formed by the vacant sites. As we can see in this example, sublattices have the separate roles of forming the crystal and superfluidity in {\it hard-core} bosonic systems. Another example is soft-core bosons on a square lattice with the nearest-neighbor repulsion $V_{\rm nn}$, where the checkerboard crystal appears like Fig. \ref{fig:superflowpath} (b) at the 1/2-filling \cite{sengupta2005}. In this case, the superflow path is formed all over the lattice for interstitials if the on-site interaction $U$ is nearly equal to $zV_{\rm nn}$, where $z$ is the coordination number. In other words, the paths allow the interstitials to hop on the checkerboard backgroud. The superflow paths are blocked in the hard-core limit $U \to \infty$. However, the checkerboard supersolid of hard-core bosons can be stabilized in the presence of the dipole-dipole interaction as mentioned above. Therefore, the present supersolid is qualitatively different from previous ones and expected to have some other mechanism.

	\begin{figure}[t]
		\includegraphics[width=5cm]{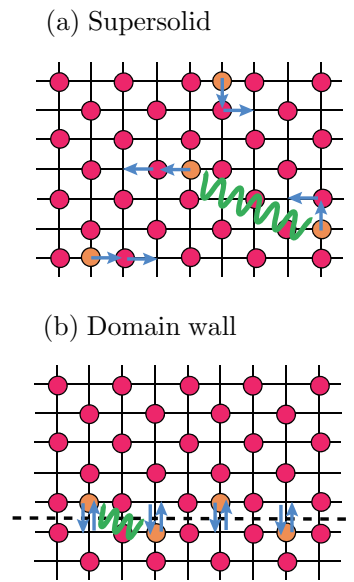}
		\caption{(Color online) The checkerboard crystal doped with interstitials. The red circles, the orange circles, and the arrows have the same meanings in Fig. \ref{fig:superflowpath}. The dashed line and the wavy line represents a domain wall and one of the repulsive interactions between interstitials respectively.  (a) Supersolid where interstitials are delocalized. (b) Domain wall where interstitials are aligned in a particular direction. In long-range interacting systems, interstitials can gain the kinetic energy by the second-order hopping process in both cases. Furthermore, repulsive interactions between interstitials appear.  }
		\label{fig:mechanism}
	\end{figure}
    
    We now discuss the checkerboard supersolid in our model. In Fig. \ref{fig:mechanism}, we show the checkerboard crystal doped with $L/2$ bosons. As illustrated in Fig. \ref{fig:mechanism} (a), supersolid can be realized by delocalization of defects\cite{andreev1969, chester1970}. If the system has only the nearest-neighbor interaction, domain wall formations like Fig. \ref{fig:mechanism} (b) are energetically more favorable than the supersolid for small hopping amplitudes, as discussed in Ref. \cite{sengupta2005}. This is due to two important features in short-range interacting systems. One is that the energetic cost for doping a single boson is independent of positions of the empty sites. The other is that interstitials forming domain walls can gain the kinetic energy by the first-order hopping process, whereas bosons can delocalize by the second-order hopping process. By these features, with doping of the small particle density $\sim 1/L$, the supersolid state(Fig. \ref{fig:mechanism} (a)) and the domain wall state(Fig. \ref{fig:mechanism} (b)) have the same energy in the zeroth-order in $t$, whereas the domain wall state has a lower energy in the first-order in $t$. Consequently, the supersolid becomes unstable to the domain wall formation.

   However, the situation should be changed when the system has the long-range interactions, because the repulsive interaction does work between interstitials. As we can see in Fig. \ref{fig:mechanism} (b), interstitials forming the domain wall feel repulsive energies with each other in the presence of the third or higher-nearest neighbor repulsion. By these repulsions, it costs higher energy than the configuration where interstitials randomly occupy empty sites like Fig. \ref{fig:mechanism} (a) even in the zeroth-order of $t$. In the domain wall case, the kinetic energy gain derived from the second-order hopping process apparently exits. Since the gain is the same order in $t$ as that in the case shown in Fig. \ref{fig:mechanism} (a), it is naively expected that configuration energy of interstitials decides everything; the system prefers the supersolid state energetically.

    In order to confirm quantitatively that the long-range interactions are relevant and change the stability of the supersolid against the domain wall, we compare the energies of Fig. \ref{fig:mechanism} (a) and (b) by the perturbative calculations. For simplicity, we consider the dipole-dipole interaction up to the fourth-nearest neighbor interactions. Adding a single boson to the checkerboard crystal costs the energy $E_{0} = zV+2zV/(\sqrt{5})^{3} -\mu $. For the domain wall configuration (b), by the repulsive interactions between interstitials and the shift of the lower half configuration, the energy is raised by $2V/(\sqrt{5})^3 - V/2^3 \sim 0.054V$ per interstitial. When we introduce the kinetic terms, we obtain the kinetic energy as $\sim -14t^2/V$ by the the second-order perturbative calculation. By these terms, the energy cost for forming the domain wall state is estimated as $E_{\rm DW} \sim E_{0} + 0.054V - 14t^2/V$ per interstitial. As shown in Ref. \cite{capogrosso2010-1}, the checkerboard supersolid phase is observed around the checkerboard Mott lobes for $t/V \agt  0.05$. When we use the small hopping parameter $t/V=0.05$ for estimation of the energy above, the zeroth energy $0.054V$ is still larger than the kinetic energy gain $14t^2/V \sim 0.035V$. Thus, we obtain that $E_{\rm DW}$ is larger than $E_{0}$; $E_{\rm DW}>E_{0}$. This is a result of higher configuration energy and lower kinetic energy gain than the case with only the nearest-neighbor repulsion. On the other hand, for the supersolid state (a), the repulsive energies between the interstitials are negligible, because the interstitials are very far away from each other in the thermodynamic limit. Hence, the zeroth term is only $E_{0}$. In the presence of the kinetic term, the energy is lowered by the second or higher-order hopping process. Thus, the energy cost of a single interstitial is $E_{\rm SS} = E_{0} - O(t^2/V)$, which is smaller than $E_{0}$; $E_{\rm SS}<E_{0}$. From these estimations, we conclude that the supersolid becomes stable against the domain wall formation owing to the effect of the repulsive interactions between interstitials. The above estimation does not change up to the presence of the eighth-nearest-neighbor repulsion, because the second-nearest interstitials interact by the ninth-nearest-neighbor repulsion which is negligibly small in the estimation above. More specifically, the third and fourth-nearest-neighbor interactions play the dominant role of increasing the energy cost of the domain wall formation. Because of the absence of these interactions, the checkerboard supersolid have not been found in systems with only the nearest neighbor repulsions and the next-nearest neighbor repulsions. Although we have discussed the interstitial-based supersolid above, the vacancy-based supersolid is also possible by the same mechanism because of the particle-hole symmetry in hard-core bosonic systems.

%\section{\label{sec:level4}Quantum Monte Carlo simulations}
	\begin{figure}[t]
		\includegraphics[width=7.3cm]{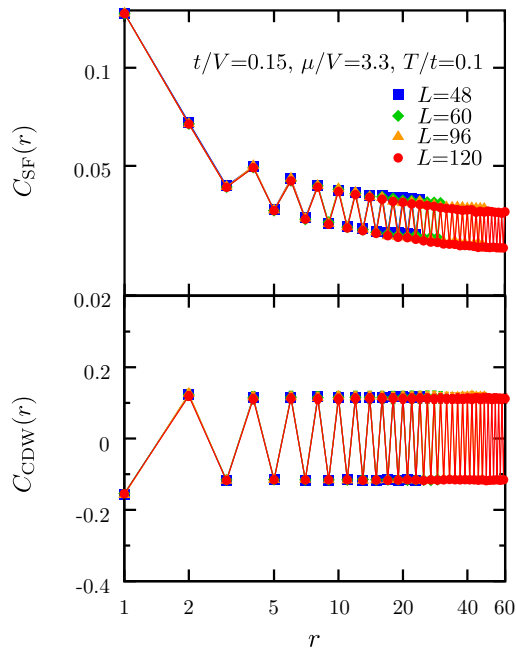}% Here is how to import EPS art
		\caption{\label{fig:correl} (Color online)  Off-diagonal correlation function $C_{\rm SF}(r)$ and diagonal correlation function $C_{\rm CDW}(r)$ as a function of the distance along the $x$-axis. $C_{\rm SF}(r)$ and $C_{\rm CDW}(r)$ are shown as logarithmic plots and semi-logarithmic plots respectively. Error bars are drawn, but most of them are smaller than the symbol size. }
	\end{figure}
 
	\begin{figure}[t]
		\includegraphics[width=7cm]{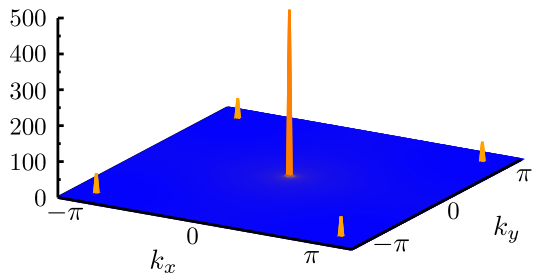}% Here is how to import EPS art
		\caption{\label{fig:ft_correl} (Color online)  Momentum distributions of bosons $n_{\rm SF}(k_{\rm x}, k_{\rm y})$ at $L=120$. The other parameters are the same as in Fig. \ref{fig:correl}. The central sharp peak is located at $\mbox{\boldmath $k$}=\mbox{\boldmath $0$}$. The four smaller peaks are at $\mbox{\boldmath $k$}=(\pi, \pi)$ and its equivalent points in the momentum space.}
	\end{figure}
 
	We next perform exact quantum Monte Carlo simulations to investigate the nature of the supersolid phase described above. The notorious difficulty in simulations of long-range interacting systems is the computational cost $O(N^2)$, which is $O(N)$ if the system has only short-range interactions. In order to overcome this difficulty, we use a new hybrid algorithm\cite{ohgoe_unpublished} of modified directed loop algorithm\cite{kato2009} and $O(N)$ Monte Carlo method\cite{fukui2009}. In our simulations, we applied the Ewald summation method\cite{deLeeuw1980} to avoid any cutoff dependence of the long-range interactions.

	In Fig. \ref{fig:correl}, we show the off-diagonal correlation function $C_{\rm SF}(\mbox{\boldmath $r$}_{ij})=\langle b_{i} b^{\dagger}_{j}\rangle$ and the diagonal correlation function $C_{\rm CDW}(\mbox{\boldmath $r$}_{ij})=\langle n_{i} n_{j}\rangle - \langle n_{i} \rangle^2$ as a function of the distance along the $x$-axis. The parameters are chosen as $(t/V, \mu/V, T/t)=(0.15, 3.3, 0.1)$ where the vacancy-based checkerboard supersolid phase appears. To observe the long-distance behavior of the correlation functions, we perform simulations up to the system size $120 \times 120$, which needs high computational cost without the $O(N)$ method. From this result, we can find that $C_{\rm SF}(r)$ shows a power-law decay, implying the presence of an off-diagonal quasi-long-range order. Simultaneously, $C_{\rm CDW}(r)$ shows the presence of a diagonal long-range order. Thus, this phase is the supersolid phase in a broad sense. As the most characteristic feature in this supersolid, we have observed a clear zig-zag pattern of the power-law decay in $C_{\rm SF}(r)$. Obviously, this pattern is a sign that the superfluid component is affected by the checkerboard-type structure. In Fig. \ref{fig:ft_correl}, we show the Fourier transformation of the off-diagonal correlation function $n_{\rm SF}(\mbox{\boldmath $k$})=1/N \sum_{i,j} C_{\rm SF}(\mbox{\boldmath $r$}_{ij}) e^{i \mbox{\boldmath $k$} \cdot \mbox{\boldmath $r$}_{ij}}$. As in the conventional superfluid, we can observe the sharp peak at $\mbox{\boldmath $k$}=\mbox{\boldmath $0$}$, which means superfluidity. In addition to this, as a result of the zig-zag pattern, we can also observe the other peaks at $\mbox{\boldmath $k$}=(\pi, \pi)$ and its equivalent points, although it is much weaker than the former one. The corresponding peaks can be measured by the time-of-flight experiment and the observation of the two types of peaks will be a clear evidence for supersolidity.

%\section{\label{sec:level5}Conclusion}
    In summary, we have revealed the novel mechanism of the supersolid of hardcore bosons with the dipole-dipole interaction. With repulsive long-range interactions, the superflow paths are not required any more for realizing supersolids, and the mechanism by the second-order hopping process of interstitials becomes relevant. Since the checkerboard supersolid is due to the second-order hopping process, the transition temperature $T_{c}$ should be scaled as $T_{c} \propto t^2/V$, and it can be very low in real experimental settings.

%\section*{ACKNOWLEDGMENTS}%*'ð''¯'é'ƁAæ"ª'ɔԍ†'ª''©'È'¢B
	The present work is financially supported by Global COE Program ``the Physical Science Frontier", Grant-in-Aid for Scientific Research (B) (22340111), Grant-in-Aid for Scientific Research on Priority Areas ``Novel States of Matter Induced by Frustration" (19052004) and Grant-in-Aid for Young Scientists (B) (21740245) from MEXT, Japan.
The simulations were performed on computers at the Supercomputer Center, Institute for Solid State Physics, University of Tokyo.

%\bibliography{supersolid2011}% Produces the bibliography via BibTeX.

\end{document}